\documentstyle[12pt,aasms4]{article}

\begin{document}

\title{BL Lac Objects and Relativistic Beaming Model}

\author{J.H. Fan}

\affil{CRAL Observatoire de Lyon, 9 Avenue Charles André, 69 563 Saint-Genis-Laval Cedex, France; and  \\
Center for Astrophysics, Guangzhou Normal University, Guangzhou 510400, China}

\begin{abstract}

 In this paper: 1. The assumption of the dependence of Doppler factor on the 
 emission frequency
 ($ \delta_{\nu} \approx \delta_{O}^{1+ \frac{1}{8} log {\frac{\nu}{\nu_{o}}}}$
 , Fan et al.  1993) is used to explain the observational differences  
 between the radio-selected BL Lac  objects(RBLs) and the X-Ray-Selected BL 
 Lac objects (XBLs): a) Hubble relation; b)  different multiwavelength 
 correlations; c) different regions in the effective spectral index  
 ( $\alpha_{RO} - \alpha_{OX}$) diagram; d) different polarization.
 The results suggest that RBLs  and XBLs are the same.
 2. From the analysis of the relation between infrared magnitude and redshift, 
 it is  proposed that the parent population of BL Lac objects should be 
 FRI radio galaxies and  FRII(G) radio galaxies showing the optical 
 spectra of a galaxy.  3. From the superluminal motion, the assumption  
 ($ \delta_{\nu} \approx \delta_{O}^{1+ \frac{1}{8} log {\frac{\nu}{\nu_{o}}}}$
  ) is confirmed. 4. Based on the relation between polarization and  
 Doppler factor (Fan et al. 1997), it is  proposed that the  $f$, ratio of 
 the beamed luminosity to the unbeamed  luminosity in the source frame 
 of OVVs/HPQs is smaller than that of BL Lac objects: 
 $f_{RBLs} \sim 6f_{FSRQs}$.

\end{abstract}

\keywords{Beaming effect -- BL Lacertae Objects (RBLs and XBLs) - OVVs/HPQs -- Superluminal Motion--Unified Scheme}

\section{Introduction}

 BL Lacertae objects are an extreme subclass of AGNs showing rapid and 
 large amplitude variability,  high and variable polarization, no or weak 
 emission features (EW $< 5\AA)$. From survey, BL  Lacertae objects can 
 be divided into radio selected BL Lac objects (RBLs) and X-ray selected 
 BL Lac  Objects (XBLs); or into quasar-like BL Lacertae objects (Q-BLs) 
 and X-ray strong BL Lacertae objects  (X-BLs) (Giommi et al. 1990) and  
 into high frequency peaked BL Lac objects (HBLs) and low frequency 
 peaked BL Lac objects (LBLs) (Padovani \& Giommi 1995; Urry \& Padovani 
 1995; Urry 1998). 

 They are some obvious differences between the two subclasses. Some authors 
 claimed that  they are the same class with XBLs having wider viewing angle 
 than do RBLs (Ghisellini \& Maraschi, 1989;  Xie et al. 1993; 
 Georganoppoulos \& Marsher 1996,1998; Fan et al. 1997a,b); Some authors 
 claimed  that there is a continuous spectral sequence from XBLs(HBLs) to 
 RBLs(LBLs) (Sambruna et al. 1996).

\section{Difference and Unity of BL Lac objects}
\subsection{The Observational Differences between RBLs and XBLs}
 1) RBLs do not fit the Hubble diagram so well as do XBLs;\\ 
 2) RBLs do not show any multiwavelength correlation as do XBLs;\\
 3) RBLs have higher radio and optical luminosities than do XBLs but they both 
    have almost the same X-ray luminosity within the error range;\\
 4) They occupy different regions in the effective spectral index diagram;\\
 5) The averaged polarization of XBLs is lower than that of RBLs with
    $\overline{\Pi_{RBLs}} > 10\% $ and $\overline{\Pi_{XBLs}} < 5\% $. \\

\subsection{Unity of RBLs and XBLs}

 The relativistic beaming model (Blandford \& Rees 1978) suggests that 
 the observed flux  has been boosted.$S^{ob.}=\delta^{p}S^{in.}$
 which gives
 $$m_{O}^{in.}=m_{O}^{ob.}+2.5\times p \times log \delta_{O}$$,
 we think that should use the intrinsic data 
 to compare the properties of RBLs and XBLs: For Hubble relation, 
 We proposed a method to estimate $\delta_{O}$ (Xie et al. 1991).
 When we use the $m_{V}^{in.}$ to discuss the Hubble relation, much 
 better results show up, and the  corrected data of RBLs fit the same 
 Hubble relation as XBLs (Xie et al. 1991; Fan et al. 1994) suggesting 
 they are a single class with RBLs being more strongly beamed than are 
 XBLs. From the  microvariability amplitude, Miller \& Noble (1996) also 
 showed that RBLs have higher  Lorentz factor than XBLs. 

 We assume that the Doppler factor satisfies  (Fan et al.  1993): 
  $$ \delta_{\nu} \approx \delta_{O}^{1+ \frac{1}{8} log {\frac{\nu}{\nu_{O}}}} $$ 
  which gives $\delta_{R} \sim \delta_{O}^{1.65}$. This  is confirmed by the 
 superluminal motion which gives  $\delta_{R} \sim \delta_{O}^{1.93 \pm0.22}$ 
 (Fan et al. 1996a,b).

 When the assumption is used on some RBLs with known optical Doppler factors, 
 we found that the corrected luminosity of RBLs are equal to those of XBLs 
 within the error range,  the corrected RBLs data show good multiwavelength 
 correlations, and the corrected spectral indices of RBLs shift to the XBLs 
 region (see Fan et al. 1993,1994;  Fan \& Xie 1996). The results suggest 
 that RBLs are more strongly beamed than XBLs  and that  RBLs have smaller 
 viewing angle than XBLs.   So, we hope to use this assumption to explain 
 the difference in their polarizations.

 Following the work of Urry \& Shafer (1984)  the observed luminosity
 consists of two parts, the beamed and the unbeamed, $S^{in}_{j}=fS_{unb}$, 
 $S^{ob}=(1+f)S_{unb}$,  and assuming  that the beamed flux  consists of 
 two proportional components ( the polarized and  the unpolarized ones),
   $$S_{jp}=\eta S_{jup}, S_{j}^{in} =S_{jp}+S_{jup}$$ we obtained that the
 observed polarization can be written as
 $$ \Pi^{ob.}={\frac {(1+f)\delta^{p}}{1+f\delta^{p}}}\Pi^{in.}       $$
where, $ \Pi^{in.}= {\frac {f\eta}{(1+f)(1+\eta)}} $. When we use the 
 maximum polarization and the  known Doppler factors (Fan et al. 1993;
 Ghisellini et al. 1993) to compare with the theoretical curve, 
 The points fit the theoretical curves (see Fan et al. 1997a).

\subsection{Parent Population of BL Lac Objects}

  Some authors think that the unified model includes  BL Lac objects and 
 FRI (Brown 1983; Urry et al. 1991; Bicknell 1994;  Padovani 1998), others 
 think that the unified model includes BL Lac objects, FRI and some FRII 
 as well (Owen et al. 1996, see also Rector 1998).   We think that the 
 unified model should include BL Lac objects, FRI and FRII(G)
 radio galaxy with galaxy spectrum  (Xie et al. 1993). 

 Recently, We used the faintest observed K magnitude to discuss the K-band 
 Hubble relation and found that the corrected magnitude of RBLs, $K^{corr.}$, 
 fit the same Hubble relation as the uncorrected X-ray selected BL Lac 
 objects (XBLs), FRI, and FRII(G)  confirming that BL lac 
 objects, FRI and FRII(G) are unified (Fan et al. 1997b, see also Fig. 1). 

\begin{figure}
\epsfxsize=15cm
$$
\epsfbox{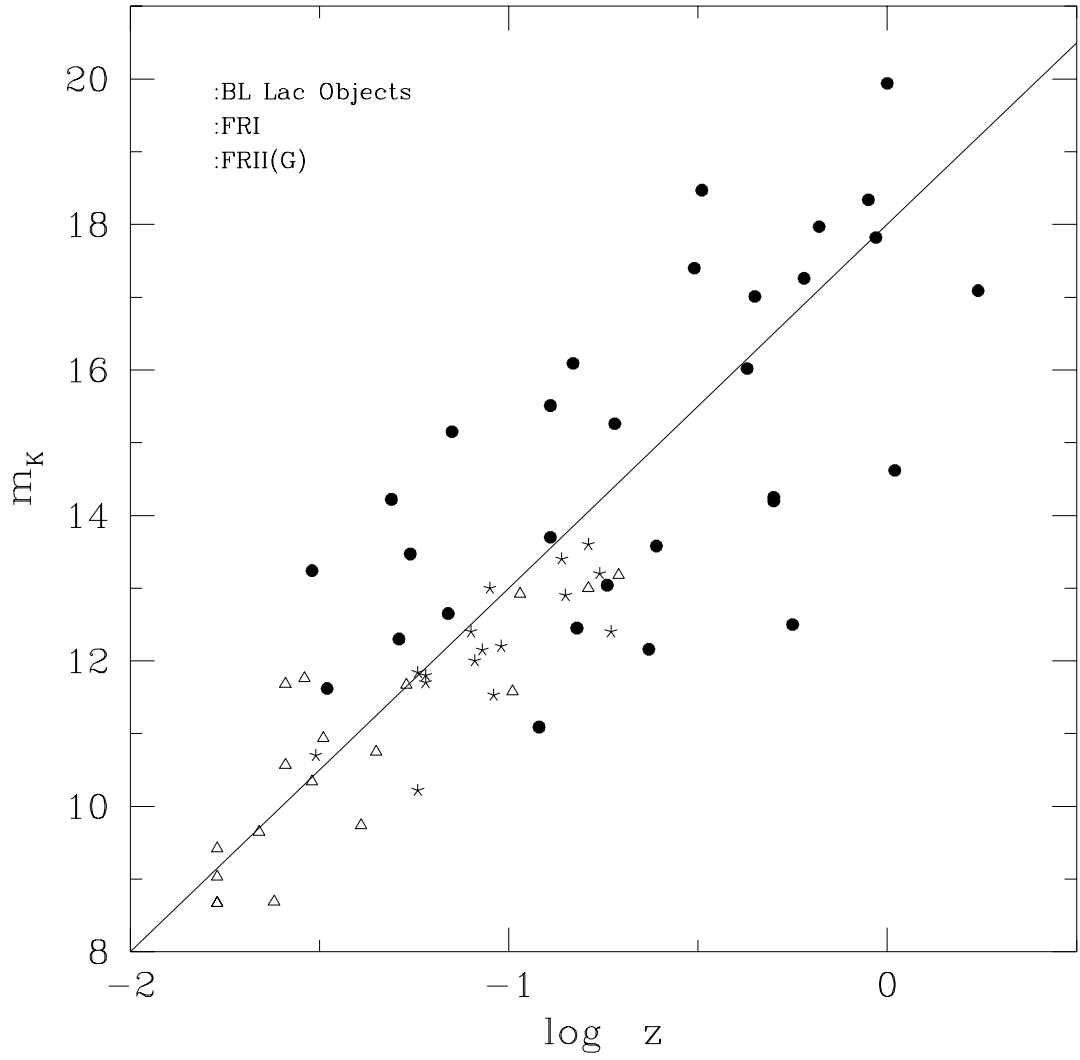}
$$

\caption{ The K-band Hubble diagram for BL Lac objects, FRI and FRII(G)}
\end{figure}

\section{RBLs and Flat Spectral Radio Quasars (FSRQs)}
 The relation between RBLs and FSRQs (HPQs/OVVs) is complex (Fan 1997a, 
 Scarpa \& Falomo 1997). We think  FSRQs have smaller $f$ than RBLs: From 
 the core-dominance parameter ($R$) expression, $ R=f\delta^{p}$, we found 
 that  FSRQs have smaller $f$'s than do RBLs (see Fan 1997b). From the 
 polarization-Doppler factor relation we can see FRSQs lie in a region 
 corresponding to smaller $f$'s as compared with RBLs (Fig. 2). So, we
 can get that  FSRQs have smaller $f$ than RBLs, and  a value of 
 $f_{RBLs} \sim 6f_{FSRQs}$ can be estimated from Fig. 2. It is clear 
 that 3C279, 3C345 and 1156 are in the region of BL Lac objects. Does 
 that mean that we should class them as RBLs? , or BL Lac objects and 
 OVVs/HPQs are originally the same with BL Lac objects showing larger $f$.

\begin{figure}
\epsfxsize=15cm
$$
\epsfbox{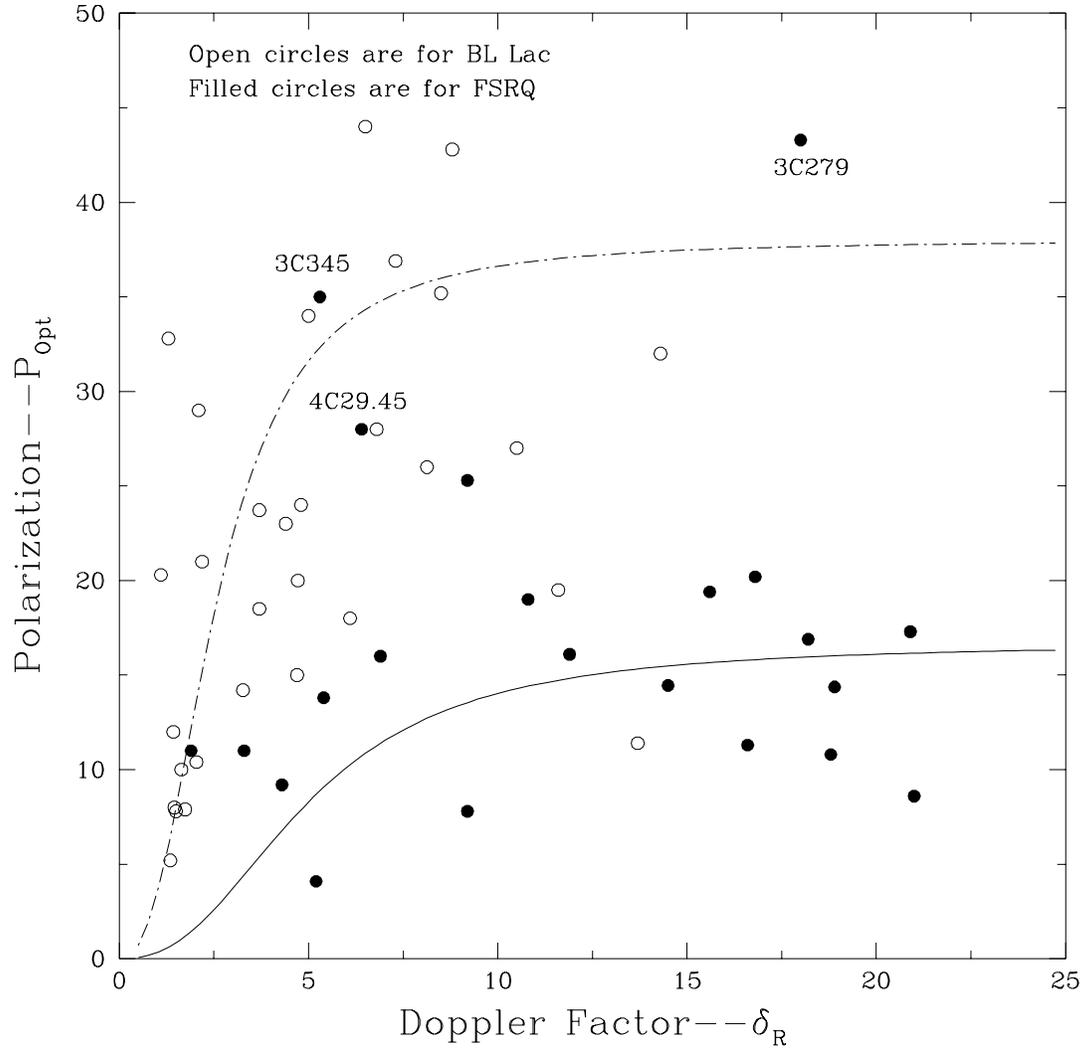}
$$

\caption{ Plot of Maximum polarization and Doppler factor for BL Lac objects 
and FSRQs}
\end{figure}

\acknowledgments
 I thank Prof. Roland Bacon, the director of l'Observatoire de Lyon for 
 his supporting my attending the conference!  This work is supported by 
 the National Pandeng Project of China

\end{document}